\documentclass[12pt]{article}
\usepackage{latexsym,amsmath}
\input amssym
\begin{document}
\renewcommand{\theequation}{\thesection.\arabic{equation}}
\def\prg#1{\medskip{\bf #1}}
\def\lra{\leftrightarrow}        \def\Ra{\Rightarrow}
\def\nin{\noindent}              \def\pd{\partial}
\def\dis{\displaystyle}          \def\dfrac{\dis\frac}
\def\grl{{GR$_\Lambda$}}         \def\vsm{\vspace{-9pt}}
\def\Lra{{\Leftrightarrow}}      \def\ads3{AdS$_3$}
\def\cs{{\scriptscriptstyle \rm CS}}
\def\Leff{\hbox{$\mit\L_{\hspace{.6pt}\rm eff}\,$}}
\def\bull{\raise.25ex\hbox{\vrule height.8ex width.8ex}}
\def\Tr{\hbox{\rm Tr\hspace{1pt}}}
\def\bF{{\bar F}}               \def\bt{{\bar\tau}}
\def\gp{($\Bbb{P}$)}            \def\tn{\tilde\nabla}
\def\tcR{\tilde{\cal R}}        \def\inn{\,\rfloor\,}
\def\bi{{\rm BI}}               \def\as{{\rm as}}

\def\D{{\Delta}}      \def\bC{{\bar C}}     \def\bT{{\bar T}}
\def\bH{{\bar H}}     \def\bL{{\bar L}}     \def\bI{{\bar I}}
\def\hO{{\hat O}}     \def\hG{{\hat G}}     \def\tG{{\tilde G}}
\def\cL{{\cal L}}     \def\cM{{\cal M }}    \def\cE{{\cal E}}
\def\cA{{\cal A}}     \def\cI{{\cal I}}     \def\cC{{\cal C}}
\def\cF{{\cal F}}     \def\hcF{\hat{\cF}}   \def\bcF{{\bar\cF}}
\def\cH{{\cal H}}     \def\hcH{\hat{\cH}}   \def\bcH{{\bar\cH}}
\def\cK{{\cal K}}     \def\hcK{\hat{\cK}}   \def\bcK{{\bar\cK}}
\def\cO{{\cal O}}     \def\hcO{\hat{\cal O}} \def\tR{{\tilde R}}
\def\cB{{\cal B}}     \def\cQ{{\cal Q}}     \def\bcL{{\bar{\cal L}}}

\def\G{\Gamma}        \def\S{\Sigma}        \def\L{{\mit\Lambda}}
\def\a{\alpha}        \def\b{\beta}         \def\g{\gamma}
\def\d{\delta}        \def\m{\mu}           \def\n{\nu}
\def\th{\theta}       \def\k{\kappa}        \def\l{\lambda}
\def\vphi{\varphi}    \def\ve{\varepsilon}  \def\p{\pi}
\def\r{\rho}          \def\Om{\Omega}       \def\om{\omega}
\def\s{\sigma}        \def\t{\tau}          \def\eps{\epsilon}
\def\ups{\upsilon}    \def\tom{{\tilde\om}} \def\bw{{\bar w}}

\def\nab{\nabla}      \def\tnab{{\tilde\nabla}}
\def\Th{\Theta}       \def\cT{{\cal T}}    \def\cS{{\cal S}}
\newcommand{\hodge}{{}^\star}
\def\nn{\nonumber}
\def\be{\begin{equation}}             \def\ee{\end{equation}}
\def\ba#1{\begin{array}{#1}}          \def\ea{\end{array}}
\def\bea{\begin{eqnarray} }           \def\eea{\end{eqnarray} }
\def\beann{\begin{eqnarray*} }        \def\eeann{\end{eqnarray*} }
\def\beal{\begin{eqalign}}            \def\eeal{\end{eqalign}}
\def\lab#1{\label{eq:#1}}             \def\eq#1{(\ref{eq:#1})}
\def\bsubeq{\begin{subequations}}     \def\esubeq{\end{subequations}}
\def\bitem{\begin{itemize}}           \def\eitem{\end{itemize}}

\title{Nonlinear electrodynamics in 3D gravity with torsion}

\author{M. Blagojevi\'c$\,^1$, B. Cvetkovi\'c$\,^1$
        and O. Mi\v skovi\'c$\,^2$
\footnote{Email addresses: {\tt mb@phy.bg.ac.yu,
          cbranislav@phy.bg.ac.yu, olivera.miskovic@ucv.cl}} \\
{\small $^1$ Institute of Physics, University of Belgrade,
P. O. Box 57, 11001 Belgrade, Serbia}\\
{\small $^2$ Instituto de F\'\i sica, Pontificia Universidad Cat{\'o}lica
      de Valpara\'\i so, Casilla 4059, Valpara\'\i so, Chile}}
\date{}
\maketitle
\begin{abstract}
We study exact solutions of nonlinear electrodynamics coupled to
three-dimensional gravity with torsion. We show that in any static
and spherically symmetric configuration, at least one component of
the electromagnetic field has to vanish. In the electric sector of
the theory, we construct an exact solution, characterized by the
azimuthal electric field. When the electromagnetic action is
modified by a topological mass term, we find two types of the
self-dual solutions.
\end{abstract}

\section{Introduction}
\setcounter{equation}{0}

Three-dimensional (3D) gravity has played an important role in our
attempts to properly understand basic features of the
gravitational dynamics \cite{1}. The study of 3D gravity led to a
number of outstanding results, among which the discovery of the
Ba\~n ados-Teitelboim-Zanelli (BTZ) black hole was of particular
importance \cite{2}. While the geometric structure of the theory
is traditionally treated as being Riemannian, Mielke and Baekler
proposed a new approach to 3D gravity \cite{3}, based on the more
general, Riemann-Cartan geometry \cite{4}. In this approach, the
geometric structure of spacetime is described by both the
curvature and the torsion, which offers an opportunity to explore
the influence of geometry on the gravitational dynamics.

Recent developments showed that 3D gravity with torsion has a
respectable dynamical structure \cite{5,6,7,8}. In particular, it
possesses a BTZ-like black hole solution, the black hole with torsion
\cite{5}, which is electrically neutral. An extension of this result to
the electrically charged sector has been studied in Refs. \cite{9,10}.
The treatment is based on Maxwell electrodynamics, and it reveals new
dynamical aspects of 3D gravity with torsion. In particular, the
structure of the new solutions, representing the azimuthal electric and
the self-dual Maxwell field, is to a large extent influenced by the
values of two different central charges that characterize the
asymptotic dynamics of the BTZ black hole with torsion.

In the 1930s, in an attempt to construct a classical theory of
charged particles with a finite self-energy, Born and Infeld
proposed a nonlinear generalization of Maxwell electrodynamics
\cite{11}. More recently, it was found that the nonlinear
Born-Infeld electrodynamics arises naturally as an effective
low-energy action in open string theory \cite{12}, which led to an
increased interest for Born-Infeld-like theories. In general
relativity with a cosmological constant, asymptotically (anti-)de
Sitter black holes were found in any dimension, and their
thermodynamic properties were discussed in a number of papers
\cite{x1}. Here, we study exact solutions for the system of {\it
nonlinear electrodynamics\/} of the Born-Infeld type coupled to 3D
gravity with torsion, generalizing thereby the analysis of
\cite{9,10}, based on {\it Maxwell electrodynamics\/}.

The layout of the paper is as follows. In section 2, we derive general
field equations describing the nonlinear electrodynamics coupled to 3D
gravity with torsion. In section 3, we use these equations to study
static and spherically symmetric field configurations. After proving a
general no-go theorem, valid for an arbitrary effective cosmological
constant $\Leff$, which states that at least one component of the
electromagnetic field has to vanish, we restrict our attention to the
electric sector of the theory. As in the Maxwell-Mielke-Baekler system
\cite{9,10}, there is no nontrivial solution corresponding to the
radial electric field. In section 4, we construct a solution generated
by the azimuthal electric field. In section 5, we consider the sector
of stationary and spherically symmetric configurations. After modifying
the electromagnetic action by a topological mass term, we find two
types of the self-dual solutions. All the solutions are defined in the
AdS sector, with $\Leff<0$. Finally, appendices contain some technical
details.

Our conventions are the same as in \cite{9,10}: the Latin indices
$(i,j,k,...)$ refer to the local Lorentz frame, the Greek indices
$(\m,\n,\l,...)$ refer to the coordinate frame, and both run over
0,1,2; the metric components in the local Lorentz frame are
$\eta_{ij}=(+,-,-)$; totally antisymmetric tensor $\ve^{ijk}$ and the
related tensor density $\ve^{\m\n\r}$ are both normalized by
$\ve^{012}=1$.

\section{Nonlinear electrodynamics coupled to 3D gravity with torsion} 
\setcounter{equation}{0}

\subsection{Geometric structure in brief}

Theory of gravity with torsion can be naturally described as a
Poincar\'e gauge theory (PGT), with an underlying spacetime structure
corresponding to Riemann-Cartan geometry \cite{4}.

Basic gravitational variables in PGT are the triad field $b^i$ and the
Lorentz connection $A^{ij}=-A^{ji}$ (1-forms). The corresponding field
strengths are the torsion and the curvature: $T^i:=db^i+A^i{_m}\wedge
b^m$, $R^{ij}:=dA^{ij}+A^i{_m}\wedge A^{mj}$ (2-forms). In 3D, we can
use the simplifying notation $A^{ij}=:-\ve^{ij}{_k}\om^k$ and
$R^{ij}=:-\ve^{ij}{_k}R^k$, which leads to
\be
T^i=db^i+\ve^i{}_{jk}\om^j\wedge b^k \, ,\qquad
R^i=d\om^i+\frac{1}{2}\,\ve^i{}_{jk}\om^j\wedge\om^k\, .   \lab{2.1}
\ee

The covariant derivative $\nabla(\om)$ acts on a general tangent-frame
spinor/tensor in accordance with its spinorial/tensorial structure;
when $X$ is a form, $\nabla X:=\nabla\wedge X$. PGT is characterized by
a useful identity:
\bsubeq\lab{2.2}
\be
\om^i\equiv\tom^i+K^i\, ,                                  \lab{2.2a}
\ee
where $\tom^i$ is the Levi-Civita (Riemannian) connection, and $K^i$
is the contortion 1-form, defined implicitly by
\be
T^i=:\ve^i{}_{mn}K^m\wedge b^n\, .                         \lab{2.2b}
\ee
Using this identity, one can express the curvature $R_i=R_i(\om)$ in
terms of its {\it Riemannian\/} piece $\tR_i=R_i(\tom)$ and the
contortion $K_i$:
\be
2R_i\equiv 2\tR_i+2\tnab K_i+\ve_{imn}K^m\wedge K^n\, .    \lab{2.2c}
\ee
\esubeq

The antisymmetry of the Lorentz connection $A^{ij}$ implies that the
geometric structure of PGT corresponds to Riemann-Cartan geometry, in
which $b^i$ is an orthonormal coframe, $g:=\eta_{ij}b^i\otimes b^j$ is
the metric of spacetime, and $\om^i$ is the Cartan connection.

In local coordinates $x^\m$, we can write $b^i=b^i{_\m}dx^\m$, the
frame $h_i=h_i{^\m}\pd_\m$ dual to $b^i$ is defined by the property
$h_i \inn b^j=h_i{^\m}b^j{_\m}=\d^j_i$, where $\inn$ is the interior
product. In what follows, we will omit the wedge product sign $\wedge$
for simplicity.

\subsection{Lagrangian}

General gravitational dynamics in Riemann-Cartan spacetime is
determined by Lagrangians which are at most quadratic in field
strengths. Omitting the quadratic terms, we arrive at the topological
\emph{Mielke-Baekler model} for 3D gravity \cite{3}:
\bsubeq\lab{2.3}
\be
I_0=\int 2ab^i R_i-\frac{\L}{3}\,\ve_{ijk}b^i b^j b^k\
    +\a_3L_\cs(\om)+\a_4 b^i T_i\, .                       \lab{2.3a}
\ee
Here, $a=1/16\pi G$ and $L_\cs(\om)$ is the Chern-Simons Lagrangian for
the Lorentz connection, $L_\cs(\om)=\om^id\om_i
+\frac{1}{3}\ve_{ijk}\om^i\om^j\om^k$. The Mielke-Baekler (MB) model is
a natural generalization of general relativity with a cosmological
constant (\grl).

The complete dynamics includes also the contribution of matter fields,
minimally coupled to gravity. In this paper, we focus our attention to
the case when matter is represented by the {\it nonlinear
electrodynamics\/}:
\be
I=I_0+I_M\, ,\qquad I_M\equiv\int L_M
                        =\int d^3 x\, b\cL_M(F^2)\, .      \lab{2.3b}
\ee
Here, the Lagrangian density $\cL_M$ is a function of the invariant
$F^2:=F_{mn}F^{mn}$, where $F=dA$ and $A$ is the electromagnetic
potential, while $b=\det(b^i{_\m}$). We also assume that in the weak
field limit, i. e. for $F^2\to 0$, $\cL_M$ reduces to the Maxwell form:
\be
\cL_M(F^2)=-\frac{1}{4}F^2+\cO(F^4)\, .                    \lab{2.3c}
\ee
\esubeq
A particularly important case of such a nonlinear electrodynamics is
the Born-Infeld theory:
\be
\cL_M=-\frac{k^2}{4}
  \left(\sqrt{1+\frac{2F^2}{k^2}}-1\right)=:\cL_{\rm BI}\,,\lab{2.4}
\ee
where $k^2$ is the Born-Infeld coupling. Any nonlinear electrodynamics
that satisfies the weak field limit \eq{2.3c} is said to be of the
Born-Infeld type.

In our exposition, we shall maximally postpone using the explicit
Born-Infeld form of $\cL_M$ for two reasons: first, the field equations
are more compact, and second, some of the conclusions are independent
of the explicit form of $\cL_M$, and consequently, more general.

\subsection{General field equations}

The variation of $I$ with respect to $b^i$ and $\om^i$ yields the
gravitational field equations:
\bea
&&2aR_i+2\a_4T_i-\L\ve_{ijk}b^jb^k=\Th_i\, ,               \nn\\
&&2\a_3R_i+2aT_i+\a_4\ve_{ijk}b^jb^k=0\, , \eea where $\Th_i:=-\d
L_M/\d b^i$ is the energy-momentum  current (2-form) of matter. In
the nondegenerate sector with $\D:=\a_3\a_4-a^2\neq 0$, these
equations can be rewritten as \bsubeq\lab{2.5} \bea
&&2T_i-p\ve_{ijk}b^jb^k=u\Th_i\, ,                         \lab{2.5a}\\
&&2R_i-q\ve_{ijk}b^jb^k=-v\Th_i\, ,                        \lab{2.5b}
\eea
\esubeq
where
\bea
&&p:=\frac{\a_3\L+\a_4 a}{\D}\, ,\qquad u:=\frac{\a_3}{\D}\, ,\nn\\
&&q:=-\frac{(\a_4)^2+a\L}{\D}\, ,\qquad  v:=\frac{a}{\D}\, .  \nn
\eea
Introducing the \emph{energy-momentum tensor} of matter by
$\cT^k{_i}:={}^*(b^k\Th_i)$, the matter current $\Th_i$ can be
expressed as follows:
\bsubeq\lab{2.6}
\bea
&&\Th_i=\frac{1}{2}\left(\cT^k{_i}\ve_{kmn}\right)b^mb^n
       =\ve_{imn}t^mb^n\, ,                                  \\
&&t^m:=-\left(\cT^m{_k}-\frac{1}{2}\d^m_k\cT\right)b^k\, ,
\eea
\esubeq
where  $\cT=\cT^k{_k}$.

Following \cite{9}, we can now simplify the field equations \eq{2.6}.
If we substitute the above $\Th_i$ into \eq{2.5a} and compare the
result with \eq{2.2b}, we find the following form of the contortion:
\bsubeq\lab{2.8}
\be
K^m=\frac{1}{2}(pb^m+ut^m)\, .                             \lab{2.8a}
\ee
After that, we can rewrite the second field equation \eq{2.5b} as
\be
2R_i=q\ve_{imn}b^mb^n-v\ve_{imn}t^m b^n\, ,                \lab{2.8b}
\ee
where the Cartan curvature $R_i$ can be conveniently calculated using
the identity \eq{2.2c}:
\be
2R_i=2\tR_i+u\tnab t_i
 +\ve_{imn}\left(\frac{p^2}{4}b^mb^n
 +\frac{up}2\,t^mb^n+\frac{u^2}4t^mt^n\right)\, .          \lab{2.8c}
\ee
\esubeq
In this form of the gravitational field equations, the role of matter
field as a source of gravity is clearly described by the 1-form $t^i$.

In the case of nonlinear electrodynamics, the energy-momentum tensor
is given by
\be
\cT^k{_i}=4\cL_M^{(1)}F^{km}F_{im}-\d^k_i\cL_M\, ,
\ee
where $\cL_M^{(1)}:=\pd\cL_M/\pd F^2$.

Varying the action with respect to $A$, one obtains the electromagnetic
field equations:
\be
d(\cL_M^{(1)}\hodge F)=0\, .                               \lab{2.10}
\ee
Equations \eq{2.8} and \eq{2.10}, together with a suitable set of
boundary conditions, define the complete dynamics of both the
gravitational and the electromagnetic field.

\section{Static and spherically symmetric configurations} 
\setcounter{equation}{0}

In order to explore basic dynamical features of the nonlinear
electrodynamics coupled to 3D gravity with torsion, we begin by looking
at \emph{static} and \emph{spherically symmetric} field configurations.
Using the Schwarz\-schild-like coordinates $x^\m=(t,r,\vphi)$, we make
the following ansatz for the triad field,
\be
b^0=Ndt\, ,\qquad b^1=L^{-1}dr\, ,\qquad b^2=Kd\vphi\, ,   \lab{3.1}
\ee
and for the Maxwell field:
 \be
F=E_rb^0b^1-Hb^1b^2+E_\vphi b^2b^0\, .                     \lab{3.2}
\ee
Here, $N,L,K$ and $E_r,H,E_\vphi$ are the unknown functions of the
radial coordinate $r$, and we have $F^2=(-E_r^2+E_\vphi^2+H^2)/2$.

\subsection{The field equations}

Using the components  of Riemannian connection  $\a,\b$ and $\g$,
defined in appendix A, the electromagnetic field equations \eq{2.10}
take the form
\bea
&&\left(E_r\cL_M^{(1)}\right)'L+\g E_r\cL_M^{(1)}=0\, ,    \nn\\
&&\left(H\cL_M^{(1)}\right)'L+\a H\cL_M^{(1)}=0\, ,        \lab{3.3}
\eea
where prime denotes the radial derivative. Note that equations \eq{3.3}
do not determine $E_\vphi$, it remains an arbitrary function of $r$.
The first integrals of these equations read:
\be
-4\cL_{M}^{(1)}E_r K=Q_{1}\, ,\qquad
-4\cL_{M}^{(1)}HN=Q_{3}\, ,                                \lab{3.4}
\ee
where $Q_1$ and $Q_3$ are constants.

To find the form of the gravitational field equations, we first
calculate the energy-momentum tensor,
\be
\cT^i{_j}=4\cL_M^{(1)} \left(
  \ba{ccc}
-\left(E_r^2+E_\vphi^2\right)-\dis\frac{\cL_M}{4\cL_M^{(1)}}& -E_\vphi H
                                                            & -E_r H\\
E_\vphi H&\left( H^2-E_r^2\right)-\dis\frac{\cL_M}{4\cL_M^{(1)}}
                                                            &E_\vphi E_r\\
E_r H&E_\vphi E_r&\left( H^2-E_\vphi^2\right)-\dis\frac{\cL_M}{4\cL_M^{(1)}}
  \ea
\right)\, ,
\ee
and the corresponding expression for $t^i$:
\bea
&&t^{0}=-\left(\frac{1}{2}\cL_M-4\cL_M^{(1)}H^2\right)b^0
        +4\cL_M^{(1)}E_\vphi Hb^1+4\cL_M^{(1)}E_rH b^2\, , \nn\\
&&t^{1}=-4\cL_M^{(1)}E_\vphi H b^0
        -\left(\frac{1}{2}\cL_M+4\cL_M^{(1)}E_\vphi^2\right)b^1
        -4\cL_M^{(1)}E_rE_\vphi b^2\, ,                    \nn\\
&&t^{2}=-4\cL_M^{(1)}E_rH b^0-4\cL_M^{(1)}E_rE_\vphi b^1
        -\left(\frac{1}{2}\cL_M+4\cL_M^{(1)}E_r^2\right)b^{2}\,.
\eea
Technical details leading to the explicit form of the gravitational
field equations \eq{2.8b} are summarized in appendix A. All the
components of these equations can be conveniently divided in two sets:
those with $(i,m,n)= (0,1,2),(2,0,1),(1,2,0)$ are called diagonal, all
the others are nondiagonal. Introducing
$$
V:=v+up/2
$$
to simplify the notation, the nondiagonal equations read:
\bsubeq\lab{3.7}
\bea
&&E_rH\left(V-\frac{u^2}{4}\cL_M\right)
  =u\left[ \frac{L}{2}\left( E_rE'_r+E_\vphi E'_\vphi+HH'\right)
   +\a E_\vphi^2\right] \, ,                               \lab{3.7a}\\
&&E_\vphi H\left(V-\frac{u^2}{4}\cL_M\right)
  =-u\a E_r E_\vphi\, ,                                    \lab{3.7b} \\
&&E_r E_\vphi\left(V-\frac{u^2}{4}\cL_M\right)
  =uH\left(LE'_\vphi+\a E_\vphi\right)\, ,                 \lab{3.7c} \\
&&HE_r\left( V-\frac{u^2}{4}\cL_M\right)
  =u\left[\frac{L}{2}\left(E_rE'_r-E_\vphi E'_\vphi+HH'\right)
   -\g E_\vphi^2\right]\, ,                                \lab{3.7d}\\
&&E_\vphi E_r\left(V-\frac{u^2}{4}\cL_M\right)
  =-u\g HE_\vphi\, ,                                       \lab{3.7e}\\
&&H E_\vphi\left(V-\frac{u^2}{4}\cL_M\right)
  =uE_r\left(LE'_\vphi+\g E_\vphi\right)\, ,               \lab{3.7f}
\eea
\esubeq
while the diagonal ones are:
\bsubeq\lab{3.8}
\bea
-2\left( \g' L+\g ^2\right)&=& 2\Leff
  +V\left[\cL_M+4\cL_M^{(1)}\left( E_r^2+E_\vphi^2\right)\right] \nn \\
 &-&\frac{u^2}{4}\cL_M\left[ \frac{1}{2}\cL_M
  +4\cL_M^{(1)}\left(E_r^2+E_\vphi^2\right)\right]
  -4uLE_r H'\cL_M^{(1)}\, ,                                \lab{3.8a}\\
-2\a \g&=&2\Leff
  +V\left[\cL_M+4\cL_M^{(1)}\left(E_r^2-H^2\right)\right]  \nn\\
 &-&\frac{u^2}{4}\cL_M\left[\frac{1}{2}\cL_M
  +4\cL_M^{(1)}\left(E_r^2-H^2\right)\right]
  +4u\left(\g -\a\right) E_r H\cL_M^{(1)}\, ,              \lab{3.8b}\\
-2\left( \a' L+\a ^{2}\right)&=& 2\Leff
  +V\left[\cL_M+4\cL_M^{(1)}\left(E_\vphi^2-H^2\right)\right]  \nn\\
 &-&\frac{u^2}{4}\cL_M\left[\frac{1}{2}\cL_M
                       +4\cL_M^{(1)}\left(E_\vphi^2-H^2\right)\right]
  +4uLHE'_r\cL_M^{(1)}\, ,                                 \lab{3.8c}
\eea
\esubeq
where $\Leff:=q-p^2/4$ is the effective cosmological constant. These
equations are invariant under the \emph{duality mapping}
\bea
&&\a\to\g\, ,\qquad \g\to\a\, ,                            \nn\\
&&E_r\to iH\, ,\qquad H\to iE_r\, ,                        \lab{3.9}
\eea
which has the same form as in \grl\ \cite{13}. Consequently, the set of
all solutions is also invariant under \eq{3.9}.

\subsection{A no-go theorem}

In the case $u=0$, the nondiagonal field equations imply
$$
E_rE_\vphi=0\,,\qquad  E_rH=0\,,\qquad E_\vphi H=0\, .
$$
This is a specific result, valid also in \grl\ \cite{13}, which
states that configurations with two nonvanishing components of the
electromagnetic field are dynamically not allowed.

In the general case $u\neq 0$, the analysis of the field equations
leads to the general no-go theorem (appendix B):
\be
E_rE_\vphi H=0\, .                                         \lab{3.10}
\ee
\bitem
\item[\bull] Static and spherically symmetric configurations with
three nonvanishing components of the electromagnetic field are
dynamically forbidden.
\eitem
The theorem is valid for any form of $\cL_M(F^2)$, and for any value of
$\Leff$. Thus, it generalizes the no-go theorem found in \cite{13},
based on Maxwell electrodynamics.

Motivated by this result, we now restrict our considerations to the
electric sector of the theory, specified by $H=0$.

\subsection{Dynamics in the electric sector} 

The condition $H=0$, which defines the electric sector, leads to a
significant simplification of the field equations. The nondiagonal
equations take the form
\bsubeq\lab{3.11}
\bea
\frac{L}{2}\left(E_rE'_r+E_\vphi E'_\vphi\right)
  +\a E_\vphi^2&=&0\, ,                                    \lab{3.11a}\\
\a E_r E_\vphi &=&0\, ,                                    \lab{3.11b}\\
E_r E_\vphi\left(V-\frac{u^2}{4}\cL_M\right) &=&0\, ,      \lab{3.11c}\\
E_r\left(LE'_\vphi+\g E_\vphi\right) &=&0\, ,              \lab{3.11d}\\
\frac{L}{2}\left( E_rE'_r-E_\vphi E'_\vphi\right)
  -\g E_\vphi^2 &=& 0\, ,                                  \lab{3.11e}
\eea
\esubeq
while the diagonal ones are
\bsubeq\lab{3.12}
\bea
-2\left( \g'L+\g ^{2}\right)&=& 2\Leff
  +V\left[ \cL_M+4\left( E_r^2+E_\vphi^2\right)\cL_M^{(1)}\right] \nn\\
 &&-\frac{u^2}{4}\cL_M\left[ \frac{1}{2}\cL_M
    +4\left(E_r^2+E_\vphi^2\right)\cL_M^{(1)}\right]\, ,   \lab{3.12a}\\
-2\a\g&=& 2\Leff
  +V\left[\cL_M+4E_r^2\cL_M^{(1)}\right]-\frac{u^2}{4}\cL_M
    \left[\frac{1}{2}\cL_M+4E_r^2\cL_M^{(1)}\right]\, ,    \lab{3.12b}\\
-2\left(\a'L+\a^2\right)&=& 2\Leff
  +V\left[\cL_M+4E_\vphi^2\cL_M^{(1)}\right]
  -\frac{u^2}{4}\cL_M\left[
  \frac{1}{2}\cL_M+4E_\vphi^2\cL_M^{(1)}\right]\, .        \lab{3.12c}
\eea
\esubeq
The analysis of these equations leads to the conclusion that the only
interesting configuration is the one defined by the azimuthal electric
field $E_\vphi$ (appendix C).

\section{Solution with azimuthal electric field} 
\setcounter{equation}{0}

Since the electromagnetic field equations do not impose any restriction
on the azimuthal electric field, it is completely determined by the
gravitational field equations.

The nondiagonal gravitational field equations \eq{3.11} are very
simple:
\bsubeq
\be
LE'_\vphi+2\a E_\vphi=0\, ,\qquad LE'_\vphi+2\g E_\vphi=0\, .
\ee
They imply $\a=\g$ and consequently
\be
N=C_1K\, , \qquad E_\vphi K^2=Q_2\, ,
\ee
\esubeq
where $C_1$ and $Q_2$ are the integration constants.

With $\a=\g$, the  set of the diagonal field equations \eq{3.12}
reduces to
\bsubeq\lab{4.2}
\bea
-\a'L &=&
  2E_\vphi^2\left(V-\frac{u^2}{4}\cL_M\right)\cL_M^{(1)}\,,\lab{4.2a}\\
-2\a^2&=&2\Leff+\left(V-\frac{u^2}{8}\cL_M\right)\cL_M\, . \lab{4.2b}
\eea
\esubeq
Moreover, since equation \eq{4.2a} is obtained as the derivative with
respect to $r$ of \eq{4.2b}, only the last equation is relevant. It is
convenient to fix the radial coordinate by choosing $K=r$, which yields
\bea
&&E_\vphi=\frac{Q_2}{r^2}\,,\qquad F^2=-\frac{2Q_2^2}{r^4}\,,\nn\\
&&L^2=-r^2\Leff
  -\frac{r^2}{2}\bcL_M\left(V-\frac{u^2}{8}\bcL_M\right)\, , \nn
\eea
where
$$
\bcL_M=\cL_M\left(\dis\frac{-2Q_2^2}{r^4}\right)\, .
$$
The expression for $L^2$ is obtained using \eq{4.2b}.

The above expressions for $N,L,K$ and $E_\vphi$ represent a complete
solution for any value of $\Leff$. In what follows, we shall restrict
our considerations to the case of negative $\Leff$,
$$
\Leff=-\frac{1}{\ell^2}\, ,
$$
which corresponds asymptotically to an AdS configuration. Using
$C_1=1/\ell$, the complete solution takes the form
\bea
&&b^0=\frac{r}{\ell}dt\, ,\qquad b^1=L^{-1}dr\, ,
  \qquad b^2=rd\vphi\, ,                                   \nn\\
&&\tom^0=-Ld\vphi\, ,\qquad \tom^1=0\, ,\qquad
  \tom^2=-\frac{L}{\ell} dt\, ,                            \nn\\
&&\om^i=\tom^i+\frac{1}{2}\left(pb^i+ut^i\right)\, ,       \nn\\
&&F=\frac{Q_2}{\ell} d\vphi dt\, ,                         \lab{4.3}
\eea
where
$$
t^0=-\frac{r\bcL_M}{2\ell}dt\, ,
  \qquad t^1=-\frac{d(r\bcL_M)}{2L}\,,
  \qquad t^2=-\frac{r\bcL_M}2d\vphi\, .
$$

\subsection{Geometric structure}

The metric of the solution \eq{4.3} has the form:
\be\lab{4.4}
ds^2=\frac{r^2}{\ell^2}dt^2-\frac{dr^2}{\dis\frac{r^2}{\ell^2}
     -\frac{r^2}2\bcL_M\left(V-\frac{u^2}{8}\bcL_M\right)}-r^2d\vphi^2\,.
\ee
The expressions for the scalar Cartan curvature and the pseudoscalar
torsion read:
\bea
&&R=-\ve^{imn}R_{imn}=-6q-\frac{v(r^3\bcL_M)'}{r^2}\, .      \nn\\
&&T=-\ve^{imn}T_{imn}=-6p+\frac{u(r^3\bar{\cL}_M)'}{r^2}\,.  \nn
\eea

The metric is free of coordinate singularities in the range of $r$ for
which $L^2(r)>0$. To examine this range, we rewrite $L^2(r)$ in the
form
\bea
&&L^2(r)=r^2\frac{u^2}{16}
         \left(\bcL_M-y^-\right)\left(\bcL_M-y^+\right)\, ,\nn\\
&&y^\pm:=\frac{4}{u^2}
         \left(V\pm\sqrt{V^2-\frac{u^2}{\ell^2}}\right)\, .\nn
\eea
As shown in section 5 of Ref. \cite{9}, the positivity of the classical
central charges $c^\mp$ ensures the conditions $V<0$ and
$V^2-{u^2}/{\ell^2}>0$. Thus, both $y^+$ and $y^-$ are real and
negative, and the metric \eq{4.4} is well-defined for those $r$, for
which:
\be
\bcL_M(r)<y^-\quad{\rm or}\quad y^+<\bcL_M(r)\, .          \nn
\ee

In further analysis, we need to know the explicit form of $\bcL_M$. In
the Born-Infeld theory, we have
$$
\bcL_\bi=-\frac{k^2}{4}
  \left(\sqrt{1-\frac{r_0^4}{r^4}}-1\right)\, ,
$$
where $r^4_0={4Q^2_2}/{k^2}$, the scalar curvature is given by
$$
R_\bi=-6q +\frac{3vk^2}{4}\left(\frac{1-r_0^4/3r^4}
                            {\sqrt{1-r_0^4/r^4}}-1\right)\,,
$$
and similarly for $T_\bi$. Thus, the metric is well-defined for $r\ge
r_0$, and both the scalar curvature and the pseudoscalar torsion are
singular at $r=r_0$.
\bitem
\item[\bull] In the Born-Infeld theory, the solution \eq{4.4} is regular
for $r>r_0$, but it does not have the black hole structure. For
$Q_2=0$, it coincides with the black hole vacuum.
\eitem

It is interesting to note that the replacement $k^2\to -k^2$ in the
Born-Infeld Lagrangian makes the solution \eq{4.4} regular in the
whole region $r>0$.

\subsection{Conserved charges}

Following the standard canonical procedure described in Ref. \cite{9},
the conserved charges defined by the Born-Infeld electrodynamics
(energy, angular momentum and the electric charge) are found to vanish:
\be
E=0\, ,\qquad M=0\, ,\qquad Q=0\, .
\ee
The vanishing of $Q$ is related to the fact that the azimuthal electric
field does not produce the radial flux, while the vanishing of $E$ and
$M$ is a consequence of the rapid asymptotic fall-off of the solution
\eq{4.3}.

\section{Solutions with self-dual electromagnetic field}
\setcounter{equation}{0}

The solution with azimuthal electric field is the only interesting
configuration in the sector of {\it static\/} and spherically symmetric
configurations (Appendix C). In this section, we discuss another exact
solution, the solution with self-dual electromagnetic field, which
belongs to the sector of {\it stationary\/} and spherically symmetric
configurations.

\subsection{The field equations}

We consider an action for the nonlinear electrodynamics, modified by a
topological (Chern-Simons) mass term:
\bsubeq
\be
I_M=\int d^3x\left(b\cL_M-\m\ve^{\m\n\r}A_\m\pd_\n A_\r\right)\,,
\ee
where $\m$ is the mass parameter. The presence of the mass term
modifies the field equations for the electromagnetic field:
\be
d{\,}\left(\cL_M^{(1)}\hodge F-\frac{1}{4}\m F\right)=0\,,\lab{5.1b}
\ee
\esubeq

Looking for {\it stationary and spherically symmetric\/} field
configurations, we adopt the following ansatz for the triad field,
\bsubeq
\be
b^0=Ndt\, ,\qquad b^1=L^{-1}dr\, ,\qquad b^2=K(d\vphi+Cdt)\, ,
\ee
and for the electromagnetic field strength:
\be
F=Eb^0b^1-Hb^1b^2\, .
\ee
\esubeq
Here, $N,L,C,K$ and $E,H$ are six unknown functions of the radial
coordinate $r$.

Now, we assume the generalized {\it self-duality\/} condition for the
electromagnetic field:
\be
E=\eps H\, , \qquad \eps^2=1\, .                           \lab{5.3}
\ee
Combining the self-duality condition with the adopted behavior of
$\cL_M$ in the weak field limit \eq{2.3c}, we obtain the following
relations:
\be
F^2=0\, ,\qquad \cL_M^{(1)}=-\frac{1}{4}\, .               \lab{5.4}
\ee
After that, the electromagnetic field equations \eq{5.1b} reduce to the
simple form
\be
d{\,}\left(\hodge F+\m F\right)=0\, ,                      \nn
\ee
which coincides with the Maxwell field equations \cite{10}.

Now, we focus our attention on the gravitational field equations. Using
again \eq{5.4}, one finds the simplified form of the electromagnetic
energy-momentum tensor:
\be
\cT^i{_j}=\left(\ba{ccc}
            E^2 & 0 & EH \\
             0  & 0 & 0  \\
            -EH & 0 & -H^2
                 \ea \right) \, ,
\ee
which turns out to be identical to the corresponding Maxwell
energy-momentum tensor. Consequently, the gravitational field equations
take the same form as in Maxwell theory coupled to 3D gravity with
torsion \cite{10}:
\bea
&&2\tR_i=\Leff\ve_{imn}b^mb^n-\tilde V(\eps)\ve_{imn}t^mb^n\, ,\\
&&\tilde V(\eps):=v+\frac{pu}{2}+\eps\frac{u}{\ell}
   \equiv \frac{1}{\D}\,\frac{1}{24\pi\ell}\,c(\eps)\, ,   \nn
\eea
where $c(\eps)$ are {\it classical central charges\/},  characterizing
the asymptotic conformal structure of 3D gravity with torsion \cite{6}.
Hence, we arrive at the following conclusion:
\bitem
\item[\bull] Consider the nonlinear electrodynamics modified by
a topological mass term, and the corresponding Maxwell theory; in the
self-dual sector of stationary and spherically symmetric configurations,
these two theories in interaction with 3D gravity
with torsion have identical field equations.
\eitem

Studying the Maxwell-Mielke-Baekler system, the authors of \cite{10}
found two self-dual solutions, one for $\m=0$ and the other for $\m\ne
0$. They represent natural generalizations of the Riemannian
Kamata-Koikawa \cite{14} and Fernando-Mansouri \cite{15} solutions,
respectively. It is now simple to conclude that these solutions are
also solutions of the nonlinear electrodynamics in interaction with 3D
gravity with torsion.  Their explicit form can be found in sections 4
and 5 of Ref. \cite{10}.

\subsection{Conserved charges}

Since the energy-momentum tensor takes the same form as in \cite{10},
the corresponding expressions for the energy and angular momentum of
the self-dual solutions remain unchanged.

The electric charge can be calculated using the Noether or the
canonical approach, with the result
\be
Q=-\int_0^{2\pi}(4b\cL_M^{(1)}F^{tr}-\m A_\vphi)d\vphi\,,
\ee
where the integral is taken over the boundary at spatial infinity. For
the specific case of the Born-Infeld theory, where the self-duality
implies $4\cL_M^{(1)}=-1$, the electric charge takes the same value as
in Maxwell electrodynamics.

Consequently, the self-dual solutions with $\m=0$ and $\mu\neq 0$ for
the Born-Infeld and Maxwell electrodynamics coupled to 3D gravity with
torsion, have the same values of the energy, angular momentum and
electric charge.

\section{Concluding remarks}

In this paper, we studied exact solutions of the nonlinear
electrodynamics in interaction with 3D gravity with torsion.

(1) We proved a general no-go theorem, stating that in any static and
spherically symmetric configuration, at least one component of the
electromagnetic field has to vanish. The theorem holds for a general
nonlinear electrodynamics, independently of its weak field limit, and
for any value of $\Leff$.

(2) Restricting our attention to the electric sector with $\Leff<0$, we
constructed a solution with azimuthal electric field, which represents
a generalization of the Cataldo solution found in \grl\ \cite{13}, and
of the related solution for the Maxwell-Mielke-Baekler system \cite{9}.
In the case of Born-Infeld electrodynamics, the geometric content of
the solution is clarified.

(3) In the self-dual sector of stationary and spherically symmetric
configurations, the field equations are seen to coincide with those for
the Maxwell-Mielke-Baekler system \cite{10}. These equations reveal a
new dynamical role of the classical central charges. Relying on the
results of \cite{10}, we found two types of the self-dual solutions,
both for $\Leff<0$, which are natural generalizations of the
Kamata-Koikawa and Fernando-Mansouri solutions \cite{14,15}, known from
\grl.

\appendix
\section*{Acknowledgements} 

This work was supported in part by the Serbian Science Foundation under
Grant No. 141036 (M.B. and B.C.), and in part by the Chilean Grant Nos.
FONDECYT 11070146 and PUCV 123.797/2007 (O.M.).

\section{The Cartan curvature} 
\setcounter{equation}{0}

In this appendix, we present some technical details regarding the
structure of the field equations \eq{2.8b} for static and spherically
symmetric configurations.

In Schwarzschild-like coordinate $(t,r,\vphi)$, the form of the static
and spherically symmetric triad field is defined in \eq{3.1}. After
calculating the Levi-Civita connection,
\bsubeq
\be
\tom^0=-\g b^2\, ,\qquad\tom^1=0\, ,\qquad\tom^2=-\a b^0\,,\nn
\ee
where $\a=LN'/N$, $\g=LK'/K$, we find the Riemannian curvature:
\bea
&&\tR_0=-(\g'L+\g^2)b^1b^2\, ,\qquad\tR_1=-\a\g b^2b^0\, , \nn\\
&&\tR_2=-(\a'L+\a^2)b^0b^1\, .                             \nn
\eea
\esubeq

In the next step, we calculate the expressions $B_i:=\ve_{imn}t^mb^n$ and
$C_i:=\ve_{imn}t^mt^n$:
\bsubeq
\bea
&&B_{0}=4E_rH\cL_M^{(1)}b^0b^1
  -\left[\cL_M+4\left(E_\vphi^2+E_r^2\right)\cL_M^{(1)}\right]b^1b^2
  +4E_\vphi H\cL_M^{(1)}b^2b^0\, ,                         \nn\\
&&B_{1}=4E_\vphi E_r\cL_M^{(1)}b^{0}b^{1}
  -4E_\vphi H\cL_M^{(1)}b^1b^2
  -\left[\cL_M+4\left( E_{r}^{2}-H^{2}\right)\cL_M^{(1)}\right]b^2b^0\,,\nn\\
&&B_{2}=-\left[\cL_M+4\left(E_\vphi^2-H^2\right)\cL_M^{(1)}\right]b^0b^1
  -4E_r H\cL_M^{(1)}b^1b^2+4E_\vphi E_r\cL_M^{(1)}b^2b^0\,.\nn
\eea
\bea
&&C_{0}=-4E_{r}H\cL_M\cL_M^{(1)}b^0b^1
  +\cL_M\left[\frac{1}{2}\cL_M
              +4\left(E_\vphi^2+E_r^2\right)\cL_M^{(1)}\right] b^1b^2
  -4HE_\vphi\cL_M\cL_M^{(1)} b^2b^0\, ,                    \nn\\
&&C_{1}=-4E_\vphi E_r\cL_M\cL_M^{(1)}b^0b^1
  +4E_\vphi H\cL_M\cL_M^{(1)}b^1b^2+\cL_M\left[\frac{1}{2}\cL_M
  +4\left(E_r^2-H^2\right)\cL_M^{(1)}\right)b^2b^0\, ,     \nn\\
&&C_{2}=\cL_M\left[\frac{1}{2}\cL_M
           +4\left(E_\vphi^2-H^2\right)\cL_M^{(1)}\right]b^0b^1
  +4E_r H\cL_M\cL_M^{(1)}b^1b^2-4E_\vphi E_r\cL_M\cL_M^{(1)}b^2b^0\,.\nn
\eea
After using the relations \eq{3.3}, we obtain the simplified form of
$\tn t_i$:
\bea
\tn t_0&=&
  -4\left[\frac{L}{2}\left(E_rE'_r+E_\vphi E'_\vphi+HH'\right)
                  +\a E_\vphi^2\right]\cL_M^{(1)}b^0b^1    \nn\\
  &&+4LH'E_r\cL_M^{(1)}b^1b^2
    +4\a E_\vphi E_r\cL_M^{(1)}b^2b^0\, ,                  \nn\\
\tn t_1&=&-4\left(LE'_\vphi+\a E_\vphi\right) H\cL_M^{(1)}b^0b^1
  +4\left(LE'_\vphi+\g E_\vphi\right)E_r\cL_M^{(1)}b^1b^2  \nn\\
  &&-4\left(\g-\a\right)HE_r\cL_M^{(1)}b^2b^0\, ,          \nn\\
\tn t_2&=&-4LHE'_r\cL_M^{(1)}\,b^0b^1
  +4\left[\frac{L}{2}\left(E_rE'_r-E_\vphi E'_\vphi+HH'\right)
          -\g E_\vphi^2\right]\cL_M^{(1)}b^1b^2            \nn\\
  &&+4\g HE_\vphi\cL_M^{(1)}b^2b^0\, .                     \nn
\eea
\esubeq
The above results completely determine the Cartan curvature \eq{2.8c},
and lead to the second gravitational field equation as displayed in
\eq{3.7} and \eq{3.8}.

\section{The proof of {\boldmath $E_rE_\vphi H =0$}} 
\setcounter{equation}{0}

In this appendix, we prove the general no-go theorem formulated in
section 3.

We begin by assuming $E_rE_\vphi H\ne 0$. Then, the consistency of
the first three and the last three nondiagonal equations in \eq{3.7}
is ensured by a single relation:
\be\
E_\vphi'L+ E_\vphi(\a+\g)=0\qquad\Rightarrow \qquad
E_\vphi AK=C_1\, ,                                         \nn
\ee
where $C_1$ is a constant. After that, the set of equations \eq{3.7}
reduces to:
\bea
&&E_r H\left(V -\frac{u^2}{4}\cL_M\right)=
  \frac{u}{2}\left[L(E_rE'_r+HH')+(\a-\g)E_\vphi^2\right]\,,\nn\\
&&E_\vphi H\left(V-\frac{u^2}{4}\cL_M\right)=-u\a E_r E_\vphi\,,\nn\\
&&E_rE_\vphi\left(V-\frac{u^2}{4}\cL_M\right)=-u\g HE_\vphi\, .\nn                                                            \lab{B2}
\eea
Now, the last two equations imply:
\be
\a E_r^2=\g H^2\, , \qquad
  V-\frac{u^2}{4}\cL_M=-u\a\frac{E_r}{H}\, ,               \nn
\ee
whereupon \eq{3.8b} reduces to
\be
\Leff+\left(\frac{u}{4}\cL_M-\a\frac{E_r}{H}\right)^2=0\,, \nn
\ee
or equivalently:
\be
\Leff+\left(\frac{V}{u}\right)^2=0\, .                     \lab{B1}
\ee
Using $V=v+up/2\,$ and the identity $ap+\a_3q+\a_4=0$, the last relation
leads to $\a_3\a_4-a^2=0$, which is in contradiction with the property
$\D\ne 0$, adopted in section 2.

This completes the proof that the configuration $E_rE_\vphi H\ne 0$ is
not allowed dynamically, which is exactly the content of the no-go
theorem.

\section{Field configurations with {\boldmath $E_r\ne 0$}}
\setcounter{equation}{0}

In this appendix, we show that static and spherically symmetric field
configurations with the radial electric field $E_r\ne 0$ either do not
exist or are physically irrelevant.

\subsection{{\boldmath $H=0$}}

Let us first assume the existence of a general electric field,
$E_rE_\vphi\ne 0$. In this case, equations \eq{3.11b} and \eq{3.11c}
imply
$$
\a=0\, ,\qquad V-\frac{u^2}{4}\cL_M=0\, ,
$$
whereupon equation \eq{3.12b} leads to \eq{B1}, which is a
contradiction. Hence, our assumption cannot be true, i.e. at least one
of the components $E_r,E_\vphi$ must vanish.

Next, we assume $E_\vphi=0$, whereupon equation \eq{3.11a} implies
$E_r$ = const. We have here a solution with constant electric field,
which is of no physical interest.

\subsection{{\boldmath $H\ne 0$}}

In this sector, we have both electric and magnetic fields, and the
duality symmetry \eq{3.9} implies that the only nontrivial case is
$E_\vphi=0$. If we limit our attention to the AdS sector and choose
$K=r$, the asymptotic behavior of $N$ and $L$ is given by
$$
N\sim \frac{r}{\ell}+\frac{\eta}{r^{n-1}}\, ,\qquad
\frac{L}{N}\sim1+\frac{\chi}{r^m}\, ,
$$
with $n,m>0$. Since the electromagnetic field is asymptotically weak,
we have $\cL_M^{(1)}\sim-1/4$.

The nondiagonal equations \eq{3.7} reduce to a single relation,
\be
E_rH\left(V-\frac{u^2}{4}\cL_M\right)
  =\frac{uL}{2}\left( E_rE'_r+HH'\right)\, ,               \lab{C1}
\ee
while the difference between \eq{3.8a} and \eq{3.8b}, combined with
\eq{C1} and \eq{3.3}, leads to:
\be
\left(\frac{Q_3^2}{Q_1^1}-\frac{N^2}{r^2}\right)\frac{H^\prime}H=
\frac{4N^{3}}{uQ_{1}r^2H}\left(\frac Nr\right)^{\prime }
-\frac{3}{2}\left(\frac{N^2}{r^2}\right)^\prime H \, .     \lab{C2}
\ee
The consistency of \eq{C2} to leading order requires $Q_1=\eps\ell
Q_3$, where $\eps^2=1$, and we have the asymptotic duality relation
$E_r\sim\eps H$, as expected. However, the consistency of \eq{C2} to
higher orders yields additionally one of the following two conditions:
\bea
&& 3n=2\, ,\qquad \chi=0\, ,                               \nn\\
&& m=n+2\, ,\qquad
  \chi=\frac{u\ell^2 Q_1^2}{4}\frac{3n-2}{n+2}\eta\, .     \nn
\eea
In both cases ($\chi=0$ and $\chi\ne 0$), the diagonal equations imply
$\D=0$, which is not allowed. Hence, asymptotically AdS solutions we
were looking for do not exist.


\end{document}